\documentclass[aps,prb,reprint,showpacs]{revtex4-1}
\usepackage[dvipdfm]{graphicx}
\usepackage{color}
\usepackage{bm}
\begin{document}

% Use the \preprint command to place your local institutional report
% number in the upper righthand corner of the title page in preprint mode.
% Multiple \preprint commands are allowed.
% Use the 'preprintnumbers' class option to override journal defaults
% to display numbers if necessary
%\preprint{}

%Title of paper
%\title{Bogoliubov excitations and stability of two-dimensional supersolids}
%\title{Critical velocity of two-dimensional supersolids}
%\title{Stability of two-dimensional supersolids}
\title{Mean-field and stability analysis of two-dimensional flowing soft-core bosons modeling a supersolid}
%\title{Mean-field and stability analysis of two-dimensional supersolids}
%\title{Critical velocity of two-dimensional soft-core bosons}
%\title{Critical velocity of two-dimensional soft-core Bose--Einstein condensates}

% repeat the \author .. \affiliation  etc. as needed
% \email, \thanks, \homepage, \altaffiliation all apply to the current
% author. Explanatory text should go in the []'s, actual e-mail
% address or url should go in the {}'s for \email and \homepage.
% Please use the appropriate macro foreach each type of information

% \affiliation command applies to all authors since the last
% \affiliation command. The \affiliation command should follow the
% other information
% \affiliation can be followed by \email, \homepage, \thanks as well.
\author{Masaya Kunimi}
\email{kunimi@vortex.c.u-tokyo.ac.jp}
\affiliation{Department of Basic Science, The University of Tokyo, Tokyo 153-8902, Japan}
\author{Yusuke Kato}
\affiliation{Department of Basic Science, The University of Tokyo, Tokyo 153-8902, Japan}

%\homepage[]{Your web page}
%\thanks{}
%\altaffiliation{}
%\affiliation{Department of Basic Science, University of Tokyo, Tokyo 153-8902}
%Collaboration name if desired (requires use of superscriptaddress
%option in \documentclass). \noaffiliation is required (may also be
%used with the \author command).
%\collaboration can be followed by \email, \homepage, \thanks as well.
%\collaboration{}
%\noaffiliation

\date{\today}

%%%%%%%%%%%%%%%%%%%%%%%%%%%%%%%%%%%%%%%%%%%%%%%%%%%%
\begin{abstract}

The soft-core boson system is one of the simplest models of supersolids, which have both off-diagonal long-range order (Bose--Einstein condensation) and diagonal long-range order (crystalline order). Although this model has been studied from various points of view, studies of the stability of current-flowing states are lacking. Solving the Gross--Pitaevskii and Bogoliubov equations, we obtain excitation spectra in superfluid, supersolid, and stripe phases. On the basis of the results of the excitation spectra, we present a  stability phase diagram that shows the region of the metastable superflow states for each phase. 

\end{abstract}
%%%%%%%%%%%%%%%%%%%%%%%%%%%%%%%%%%%%%%%%%%%%%%%%%%%%

% insert suggested PACS numbers in braces on next line
%%%%%%%%%%%%%%%%%%%%%%%%%%%%%%%%%%%%%%%%%%%%%%%%%%%%
%%%%%%%%%%%%%%%%%%%%%%%%%%%%%%%%%%%%%%%%%%%%%%%%%%%%
% insert suggested keywords - APS authors don't need to do this
%\keywords{}
%\maketitle must follow title, authors, abstract, \pacs, and \keywords
\maketitle

%%%%%%%%%%%%%%%%%%%%%%%%%%%%%%%%%%%%%%%%%%%%%%%%%%%%%%%%%%%%%%

{\it Introduction.} A supersolid is a quantum phase that has both superfluidity and solidity. After it was demonstrated in some seminal works\cite{Andreev1969,Chester1970,Leggett1970} that this intriguing state can be realized in quantum solids, much research has been done. Recently, Kim and Chan\cite{Kim2004} presented the possibility of a supersolid in solid ${}^4$He. They reported experimental results of non-classical rotational inertia (NCRI) in solid ${}^4$He. While there have been many experimental and theoretical studies made\cite{Balibar2010,Prokofev2007}, there still remains controversy over the origin of NCRI in solid ${}^4$He. Ultra-cold atomic gases have also become a field for the study of supersolids. Many theoretical works support the existence of a supersolid phase in systems with strong long-range interactions such as dipole-dipole\cite{Goral2002} or van der Waals interactions\cite{Henkel2010}. Recently, Bose--Einstein condensates (BEC) of ${}^{52}$Cr\cite{Griesmaier2005}, ${}^{164}$Dy\cite{Lu2011}, and ${}^{168}$Er\cite{Aikawa2012}, which have large magnetic moments, have been realized experimentally. Owing to the high controllability of experimental conditions, cold atomic gases may provide insights into the nature of supersolids.

In the many previous studies, the properties of equilibrium states of supersolids have been investigated. However, dynamical stability of a finite superflow state can not be obtained by calculating equilibrium states in the absence of a current\cite{Fisher1973}.We need to investigate the stability of a current-carrying state that has both off-diagonal long-range order (ODLRO) and diagonal long-range order (DLRO). 

The stability of condensates can be investigated by calculating the excitation spectra. In many cases, the low-energy excitations determine the critical velocity of superfluids. For example, the Landau critical velocity\cite{Landau1941} is determined by the excitation energy, and the critical velocity of a condensate in a moving optical lattice can be calculated from the excitation spectrum\cite{Wu2001}. The excitation spectrum can be not only theoretically calculated but also experimentally observed in neutron scattering experiments for ${}^4$He\cite{Donnelly1981} or Bragg spectroscopy of cold atoms\cite{Steinhauer2002}. One of the most striking properties of superfluids is the existence of a critical velocity, and determining the critical velocity of a supersolid is an important problem.
To determine the supersolid critical velocity, we use a simple continuum model of supersolids that was proposed by Pomeau and Rica\cite{Pomeau1994}. Using the Gross--Pitaevskii (GP) equation with a finite-range interaction (soft-core interaction), they showed that the ground state exhibits ODLRO and DLRO at a sufficiently high density of the condensate. Although this model uses a simplified interaction potential for the purposes of manageability, it is suitable for investigating supersolidity. In fact, the properties of bosons with finite-range interactions have been studied in various contexts\cite{Josserand2007,Aftalion2007,Sepulveda2010,Watanabe2012,Saccani2011,Saccani2012}. 

In this paper, we present a phase diagram of metastable superflow states for each phase(we call this the ``stability phase diagram'' in the following) of two-dimensional soft-core bosons at zero temperature by solving the GP and the Bogoliubov equations\cite{Bogoliubov1947}. Although a similar analysis has been done for a lattice system\cite{Danshita2010}, a continuum system has not yet been studied. We find that three phases are stable against the superflow: a superfluid phase, a supersolid phase, and a stripe phase.

%%%%%%%%%%%%%%%%%%%%%%%%%%%%%%%%%%%%%%%%%%%%%%%%%%%%%%%%%%%%%%

{\it Model.} We use the two-dimensional GP equation with a finite range interaction\cite{Pomeau1994,Josserand2007,Aftalion2007,Sepulveda2010,Watanabe2012}:
\begin{eqnarray}
-\frac{\hbar^2}{2m}\nabla^2\Psi(\bm{r})\hspace{-0.1em}+\hspace{-0.4em}\int \hspace{-0.3em}d\bm{r}'V(\bm{r}-\bm{r}')|\Psi(\bm{r}')|^2\Psi(\bm{r})\hspace{-0.1em}=\hspace{-0.1em}\mu\Psi(\bm{r}),\label{eq:GP_equation}
\end{eqnarray}
where $\Psi(\bm{r})$ is the condensate wave function, $m$ is the atomic mass, $V(\bm{r}-\bm{r}')\equiv V_0\theta(a-|\bm{r}-\bm{r}'|)$ is the two-body interaction, $V_0$ is a positive constant, $a$ is the interaction range, and $\theta(x)$ denotes the Heaviside step function. The chemical potential $\mu$ is determined by the total particle number $N$. The interaction strength of this system can be measured by a dimensionless parameter\cite{Sepulveda2010} $g\equiv \pi n_0ma^4V_0/\hbar^2$, where $n_0$ is the mean-particle density. We use $g$ as a control parameter. We assume that the solution of eq.~(\ref{eq:GP_equation}) is a plane wave or has a crystalline order that can be written by the Bloch wave function\cite{note1}:
\begin{eqnarray}
\Psi(\bm{r})=e^{i\bm{q}\cdot\bm{r}}\sum_{\bm{G}}C_{\bm{q}+\bm{G}}e^{i\bm{G}\cdot\bm{r}},\label{eq:Bloch_wave_function_for_condensate_wave_function}
\end{eqnarray}
where $\hbar\bm{q}/m$ is the velocity of the condensate, $\bm{G}$ is the reciprocal lattice vector, and $C_{\bm{q}+\bm{G}}$ is an expansion coefficient. We calculate three crystalline structures: a triangular lattice, a square lattice, and a stripe structure. We optimize the period and shape of these structures to minimize the energy per particle under given parameters $\bm{q}$ and $g$. In order to obtain the excitation spectrum, we solve the Bogoliubov equation:
\begin{widetext}
\begin{eqnarray}
\epsilon u(\bm{r})&=&K u(\bm{r})+\int d\bm{r}'V(\bm{r}-\bm{r}')\left[\Psi^{\ast}(\bm{r}')\Psi(\bm{r})u(\bm{r}')-\Psi(\bm{r}')\Psi(\bm{r})v(\bm{r}')\right],\label{eq:Bogoliubov_u}\\
\epsilon v(\bm{r})&=&-Kv(\bm{r})-\int d\bm{r}'V(\bm{r}-\bm{r}')\left[\Psi(\bm{r}')\Psi^{\ast}(\bm{r})v(\bm{r}')-\Psi^{\ast}(\bm{r}')\Psi^{\ast}(\bm{r})u(\bm{r}')\right],\label{eq:Bogoliubov_v}\\
K&\equiv &-\frac{\hbar^2}{2m}\nabla^2-\mu+\int d\bm{r}'V(\bm{r}-\bm{r}')|\Psi(\bm{r}')|^2,
\end{eqnarray}
\end{widetext}
where $\epsilon$ is the excitation energy and $u(\bm{r})$ and $v(\bm{r})$ are excitation wave functions. Using the Bloch theorem, we can expand $u(\bm{r})$ and $v(\bm{r})$ in terms of the reciprocal lattice vector:
\begin{eqnarray}
u_{\bm{k}, n}(\bm{r})&=&e^{i\bm{q}\cdot\bm{r}}\sum_{\bm{G}}A_{\bm{k}+\bm{G}, n}e^{i(\bm{k}+\bm{G})\cdot\bm{r}},\label{eq:expansion_of_u}\\
v_{\bm{k}, n}(\bm{r})&=&e^{-i\bm{q}\cdot\bm{r}}\sum_{\bm{G}}B_{\bm{k}+\bm{G}, n}e^{i(\bm{k}-\bm{G})\cdot\bm{r}},\label{eq:expansion_of_v}
\end{eqnarray}
where $\bm{k}$ is the wave number vector of the excitations, $n$ is the band index, and $A_{\bm{k}+\bm{G}, n}$ and $B_{\bm{k}+\bm{G}, n}$ are expansion coefficients. Solving the GP and the Bogoliubov equations for the four assumed states, we obtain the excitation energy and local stability of each phase\cite{supplemental}. We restrict the number of expansion coefficients to 29, 73, and 89 for the stripe structure, triangular lattice, and square lattice around the origin of the reciprocal lattice space, respectively. We have checked the convergence of the present results by comparing the results with 27, 61, and 81 expansion coefficients for the stripe structure, triangular lattice, and square lattice, respectively.

%%%%%%%%%%%%%%%%%%%%%%%%%%%%%%%%%%%%%%%%%%%%%%%%%%%%%%%%%%%%%%

\begin{figure}[h]
\includegraphics[width=7.0cm,clip]{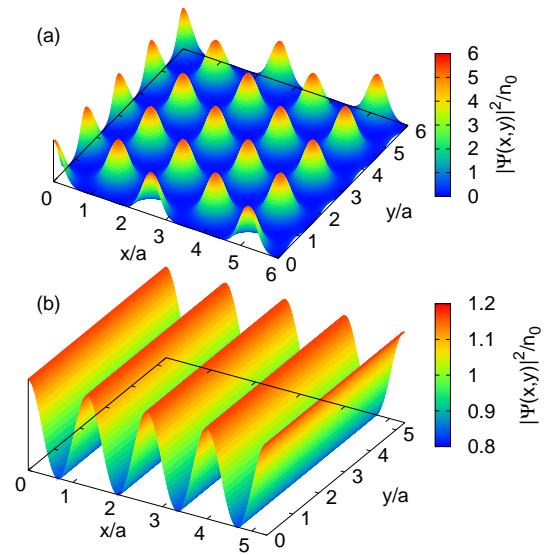}
\caption{(Color online) (a) Density profiles of the SS phase for $(g, q a)=(45, 0)$ and (b) the stripe phase for $(g, q a)=(37.5, 1.05)$.}
\label{fig:density_profile_for_g=40}
\end{figure}% 

\begin{figure}[h]
\includegraphics[width=7.5cm,clip]{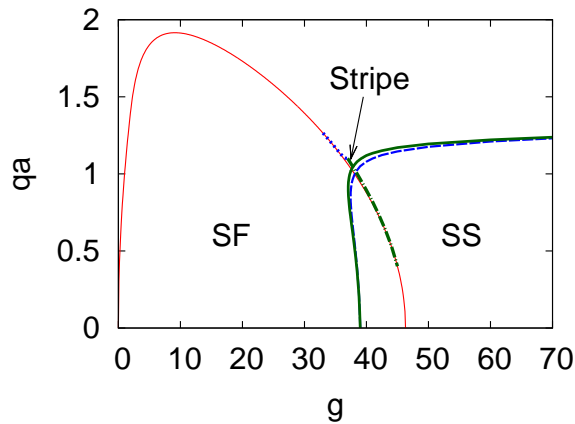}
\caption{(Color online) Stability phase diagram in the $(g, q a)$ plane. The thin solid red , dashed blue, thick solid green, dotted blue, and dashed-dotted green lines represent the Landau critical velocity of the SF phase, Landau instability(LI) line of the SS phase, dynamical instability(DI) line of the SS phase, DI at the long wavelength limit line of the stripe phase, and DI at finite $k$ line of the stripe phase, respectively. The SF phase is metastable in the region surrounded by the thin solid red line and the line with $q a=0$. The SS phase is metastable on the right side of the dashed blue line. There is no metastable stationary solution in the region not surrounded by any lines. Since we can not see the metastable region of the stripe phase in the present plot range, the magnified figure is shown in Fig.~\ref{fig:zoomed_stability_phase_diagram_1}. }
\label{fig:stability_phase_diagram_1}
\end{figure}%

\begin{figure}[h]
\includegraphics[width=7.5cm,clip]{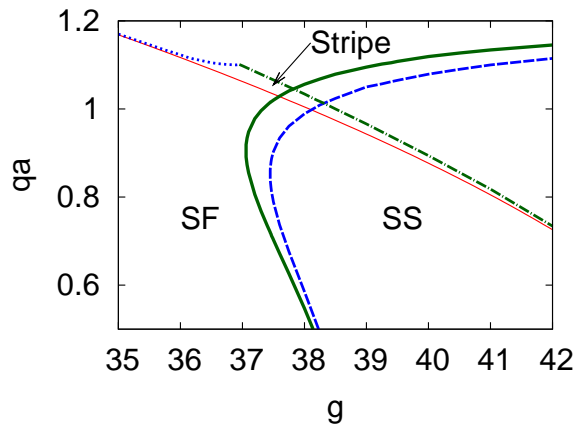}
\caption{(Color online) Magnified view of the region near the stripe phase of Fig.~\ref{fig:stability_phase_diagram_1}. The stripe phase is metastable between the thin solid red line and the dotted blue and dashed-dotted green lines. }
\label{fig:zoomed_stability_phase_diagram_1}
\end{figure}%

%%%%%%%%%%%%%%%%%%%%%%%%%%%%%%%%%%%%%%%%%%%%%%%%%%%%%%%%%%%%%%

{\it Results.}
First, we show the ground state, which corresponds to the case of $\bm{q}=\bm{0}$. In our calculation, we find that a square lattice structure is always dynamically unstable and the stripe phase is not realized at $\bm{q}=\bm{0}$. Henceforth, we use the term ``supersolid(SS) phase" only for a triangular lattice. The definitions of these phases are summarized as follows: the superfluid(SF) phase is the state in which the density of the condensate is uniform, the SS phase has a triangular lattice structure, and the stripe phase has a one-dimensional periodic structure. A typical density profile of the SS phase is shown in Fig.~\ref{fig:density_profile_for_g=40}~(a). Comparing the energies of the stable stationary solutions for SF and SS phases, we find that the SF (SS) phase has a lower energy than SS (SF) phase at $g<g_{\rm c}\simeq 39.49$ $(g>g_{\rm c})$. However, the chemical potentials of the SF and SS phases are not equal at $g=g_{\rm c}$. This implies that an inhomogeneous phase ( = coexistence phase) is realized as the ground state in the vicinity of $g_{\rm c}$. We determine the coexistence region as $38.44\le g\le 40.98$; the SF phase is realized for $0<g\le 38.44$ and the SS phase for $40.98\le g$. The result of the transition point is consistent with that of Ref.~\onlinecite{Watanabe2012}\cite{note2}.

Next, we consider the current-flowing states with  $\bm{q}\not=\bm{0}$. The current is assumed to be parallel to the $x$ direction~($\bm{q}\equiv (q, 0), q>0$). We do not consider coexistence phase but single phases in the following (we will discuss this point later). Figures~\ref{fig:stability_phase_diagram_1} and \ref{fig:zoomed_stability_phase_diagram_1} show the stability phase diagram that represents the region of single phase of metastable superflow states in the $(g, q a)$ plane. 

In the SF phase, the condensate wave function and the chemical potential are given by $\Psi(\bm{r})=\sqrt{n_0}e^{i\bm{q}\cdot\bm{r}}$ and $\mu = \pi n_0V_0a^2+\hbar^2\bm{q}^2/(2m)$. Substituting these expressions into the Bogoliubov equation, we can obtain the analytical expression of the excitation spectrum in the SF phase:
\begin{eqnarray}
\epsilon_{\bm k}=\frac{\hbar^2\bm{q}\cdot\bm{k}}{m}+\sqrt{\frac{\hbar^2k^2}{2m}\left[\frac{\hbar^2k^2}{2m}+4\pi n_0V_0a^2\frac{J_1(ka)}{ka}\right]},
\end{eqnarray}
where $J_1(x)$ is a Bessel function. In the case of $\bm{q}=\bm{0}$, this spectrum has a roton minimum when $g\ge 15.81$ and the roton gap vanishes at $g\simeq 46.30$. The metastable region of the SF phase is bounded by the thin solid red lines in Figs.~\ref{fig:stability_phase_diagram_1} and \ref{fig:zoomed_stability_phase_diagram_1}, which represent the  Landau critical velocity\cite{Landau1941}. Outside of this boundary, SF phase is unstable against the superflow.

\begin{figure}[h]
\includegraphics[width=7.0cm,clip]{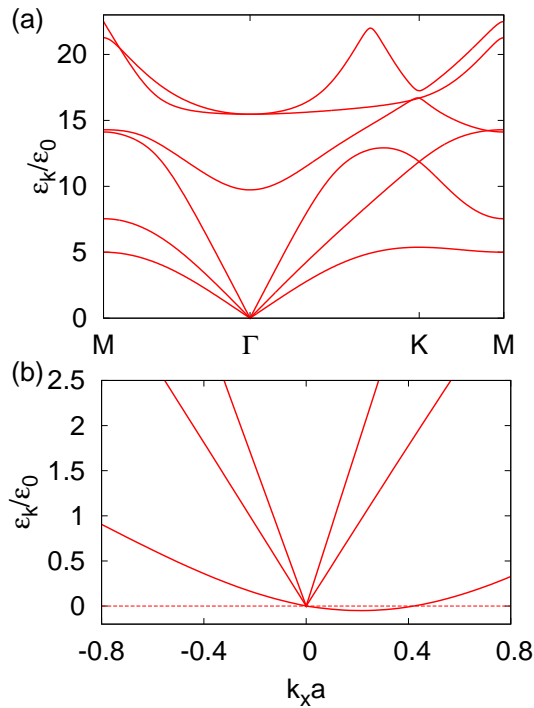}
\caption{(Color online) (a) Excitation spectrum of the SS phase for $(g, q a)=(45, 0)$. $\epsilon_{0}\equiv \hbar^2/ma^2$ is the energy unit. The coordinates of $\Gamma$, $M$, and $K$ points are given by $(0, 0)$, $(2\pi /\sqrt{3}\lambda, 0)$, and $(2\pi /\sqrt{3}\lambda, 2\pi /3\lambda)$, where $\lambda$ is a lattice constant. Similar results have been shown in Ref.~\onlinecite{Watanabe2012}. (b) Excitation spectrum of the SS phase along $k_y=0$ for $(g, q a)=(45, 1.17)$.}
\label{fig:excitation_spectrum_for_g=40}
\end{figure}

The metastability of the SS phase can be judged from the excitation spectra $\epsilon_{\bm{k}, n}$, which are obtained by numerical calculations. Figure~\ref{fig:excitation_spectrum_for_g=40} (a) shows the typical excitation spectrum of the SS phase at $\bm{q}=\bm{0}$. There are three branches of gapless excitation in the long wavelength limit. These modes are Nambu--Goldstone modes: one is a Bogoliubov mode due to spontaneous $U(1)$ symmetry breaking and the others are transverse and longitudinal phonon modes due to spontaneous breaking of translational symmetry . In the absence of a supercurrent, the excitation spectrum in the SS phase of this model and that of more general systems were studied in Refs.~\onlinecite{Watanabe2012} and \onlinecite{Son2005}, respectively. We note that the lowest branch in the SS phase is the Bogoliubov mode, which causes instabilities in the SS phase as described later.

The metastable region of the SS phase is shown as the region on the right side of the dashed blue line or thick solid green line in Fig.~\ref{fig:zoomed_stability_phase_diagram_1}\cite{note3}. Neither stable nor metastable stationary solution of the SS phase exists on the left side of these lines. The dashed blue and thick solid green line represent the Landau instability(LI) and dynamical instability(DI) lines of the SS phase, respectively; LI and DI, respectively, denote the instabilities caused by the excitation whose spectra $(\epsilon_{\bm{k}, n})$ have negative real part and nonzero imaginary part. Figure 4 (b) shows that the Bogoliubov mode that has a negative real part, which destabilizes the SS phase. These instabilities (LI and DI) occur at the long wavelength limit in the SS phase.

\begin{figure}[h]
\includegraphics[width=7.5cm,clip]{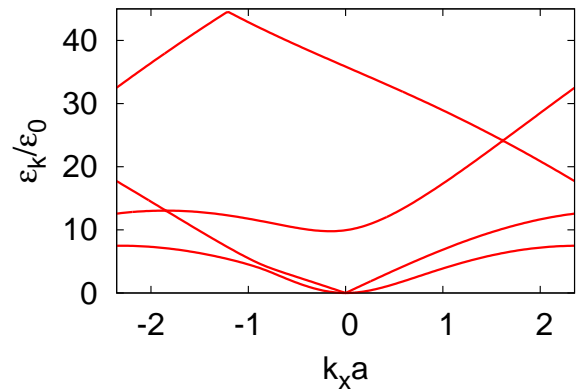}
\caption{(Color online) Excitation spectrum along $k_y=0$ in the stripe phase for $(g, q a)=(37.5, 1.05)$.}
\label{fig:excitation_stripe_1}
\end{figure}

The metastable stripe phase exists in the region surrounded by the thin solid red line and the dotted blue and dashed-dotted green lines in Fig.~\ref{fig:zoomed_stability_phase_diagram_1}. Typical density profile of the metastable stripe phase is shown in Fig.~\ref{fig:density_profile_for_g=40}~(b). The dotted blue and dashed-dotted green lines in Fig.~\ref{fig:zoomed_stability_phase_diagram_1} represent DI at the long-wavelength limit and DI at finite $k$, respectively. Both of these instabilities occur in the longitudinal phonon mode unlike in the case of the SS phase. The mechanism through which the stripe phase is generated was discussed in Refs.~\onlinecite{Pitaevskii1984,Ancilotto2005,Kunimi2011}; namely, a negative excitation mode occurs at finite $k$, leading to instability and the growth of a one-dimensional periodic structure along the flow direction. This is the reason why the stripe phase exists above the LI line of the SF phase. The period of the stripe phase is determined by the wave number that yields the roton minimum, as expected from Ref.~\onlinecite{Pitaevskii1984}. The excitation spectrum in the stripe phase is plotted in Fig.~\ref{fig:excitation_stripe_1}. In contrast to the SS phase, there are two gapless modes in the long wavelength limit. The lowest gapless mode is the longitudinal phonon mode and the other is the Bogoliubov mode. Since translational symmetry is broken in only one direction, the only two gapless modes exist in the long wavelength limit. 

Although we have focused on the metastable states in the form of Eq.~(\ref{eq:Bloch_wave_function_for_condensate_wave_function}), there are many branches of the metastable states that cannot be written by Eq.~(\ref{eq:Bloch_wave_function_for_condensate_wave_function}) such as a coexistence phase for nonzero $\bm{q}$. Further, stationary solutions including a point defect or a complex network of defects have been reported in Ref.~\onlinecite{Sepulveda2010}. In reality, which states are realized among metastable states depends on the initial condition and experimental procedures, for example, how the velocity of a container is developed from zero as a function of time. In order to determine the final state, we need to calculate a real-time and real-space dynamics with a specified protocol. This is a future work. 
Our results, on the other hand, implies that we can reach long-lived superfluid, stripe and supersolid phases only when the values of $(g,qa)$ correspond to the regions shown in Figs.~2 and 3 for respective phases. Particularly, we see that realization of a stripe phase requires a fine tuning of the parameter $(g,qa)$ in the present model. 

%%%%%%%%%%%%%%%%%%%%%%%%%%%%%%%%%%%%%%%%%%%%%

%%%%%%%%%%%%%%%%%%%%%%%%%%%%%%%%%%%%%%%%%%%%%%%%%%%%%%%%%%%%%%
%\newpage
{\it Summary and discussion.} 
In summary, we investigated the nature of the two-dimensional soft-core bosons at zero temperature by solving the GP and Bogoliubov equations. The superfluid, supersolid, and coexistence phases appear as the ground states. We presented the stability phase diagram, which represents the region of the homogeneous metastable states. The metastable superfluid, supersolid, and stripe phases are realized.

The problem of the presence of impurities or obstacles in the supersolid phase is a consideration for future work. The authors of Ref.~\onlinecite{Pomeau1994} concluded that the superfluidity of a supersolid phase in the presence of an obstacle vanishes. However, in our calculation, a supersolid phase still exhibits a supercurrent in the presence of an obstacle\cite{Anagama2012}. This discrepancy could be solved from the viewpoint of the stability analysis.

%%%%%%%%%%%%%%%%%%%%%%%%%%%%%%%%%%%%%%%%%%%%%%%%%%%%%%%%%%%%%%
%{\it Acknowledgment}\\
%\begin{acknowledgments}

We thank E. Arahata, H. Watanabe, D. A. Takahashi, and G. Anagama for useful discussions.
M. K. acknowledges the support of a Grant-in-Aid for JSPS Fellows (239376).
This work is supported by KAKENHI (21540352) and (24543061) from JSPS and (20029007) from MEXT in Japan. 
%\end{acknowledgments}

%\bibliography{basename of .bib file}

\begin{thebibliography}{99}
\bibitem{Andreev1969}
A. F. Andreev and I. M. Lifshitz, Sov. Phys. JETP {\bf 29}, 1107 (1969).
\bibitem{Chester1970}
G. V. Chester, Phys. Rev. A {\bf 2}, 256 (1970).
\bibitem{Leggett1970}
A. J. Leggett, Phys. Rev. Lett. {\bf 25}, 1543 (1970).
\bibitem{Kim2004}
E. Kim and M. H. W. Chan, Nature (London) {\bf 427}, 225 (2004); Science {\bf 305}, 1941 (2004).
\bibitem{Balibar2010}
S. Balibar, Nature (London) {\bf 464}, 176 (2010).
\bibitem{Prokofev2007}
N. Prokof'ev, Adv. Phys. {\bf 56}, 381 (2007).
\bibitem{Goral2002}
K. G\'{o}ral, L. Santos, and M. Lewenstein, Phys. Rev. Lett. {\bf 88}, 170406 (2002); B. Capogrosso-Sansone, C. Trefzger, M. Lewenstein, P. Zoller, and G. Pupillo, {\it ibid}. {\bf 104}, 125301 (2010); L. Pollet, J. D. Picon, H. P. B\"{u}chler, and M. Troyer, {\it ibid}. {\bf 104}, 125302 (2010).
\bibitem{Henkel2010}
N. Henkel, R. Nath, and T. Pohl, Phys. Rev. Lett. {\bf 104}, 195302 (2010); F. Cinti, P. Jain, M. Boninsegni, A. Micheli, P. Zoller, and G. Pupillo, {\it ibid}. {\bf 105}, 135301 (2010).
\bibitem{Griesmaier2005}
A. Griesmaier, J. Werner, S. Hensler, J. Stuhler, and T. Pfau, Phys. Rev. Lett. {\bf 94}, 160401 (2005).
\bibitem{Lu2011}
M. Lu, N. Q. Burdick, S. H. Youn, and B. L. Lev, Phys. Rev. Lett. {\bf 107}, 190401 (2011).
\bibitem{Aikawa2012}
K. Aikawa, A. Frisch, M. Mark, S. Baier, A. Rietzler, R. Grimm, and F. Ferlaino, Phys. Rev. Lett. {\bf 108}, 210401 (2012).
\bibitem{Fisher1973}
M. E. Fisher, M. N. Barber, and D. Jasnow, Phys. Rev. A, {\bf 8}, 1111 (1973).
\bibitem{Landau1941}
L. D. Landau, J. Phys. (USSR) {\bf 5}, 71 (1941).
\bibitem{Wu2001}
B. Wu and Q. Niu, Phys. Rev. A, {\bf 64}, 061603(R) (2001); New. J. Phys. {\bf 5}, 104 (2003); M. Kr\"{a}mer, C. Menotti, L. Pitaevskii, and S. Stringari, Eur. Phys. J. D {\bf 27}, 247 (2003).
\bibitem{Donnelly1981}
R. J. Donnelly, J. A. Donnelly, and R. N. Hills, J. Low. Phys. {\bf 44}, 471 (1981).
\bibitem{Steinhauer2002}
J. Steinhauer, R. Ozeri, N. Katz, and N. Davidson, Phys. Rev. Lett. {\bf 88}, 120407 (2002).
\bibitem{Pomeau1994}
Y. Pomeau and S. Rica, Phys. Rev. Lett. {\bf 72}, 2426 (1994).
\bibitem{Josserand2007}
C. Josserand, Y. Pomeau, and S. Rica, Phys. Rev. Lett. {\bf 98}, 195301 (2007).
\bibitem{Aftalion2007}
A. Aftalion, X. Blanc, and R. L. Jerrard, Phys. Rev. Lett. {\bf 99}, 135301 (2007).
\bibitem{Sepulveda2010}
N. Sep\'{u}lveda, C. Josserand, and S. Rica, Eur. Phys. J. B {\bf 78}, 439 (2010).
\bibitem{Watanabe2012}
H. Watanabe and T. Brauner, Phys. Rev. D, {\bf 85}, 085010 (2012).
\bibitem{Saccani2011}
S. Saccani, S. Moroni, and M. Boninsegni, Phys. Rev. B, {\bf 83}, 092506 (2011).
\bibitem{Saccani2012}
S. Saccani, S. Moroni, and M. Boninsegni, Phys. Rev. Lett. {\bf 108}, 175301 (2012).
\bibitem{Bogoliubov1947}
N. N. Bogoliubov, J. Phys. USSR {\bf 11}, 23 (1947).
\bibitem{Danshita2010}
I. Danshita and D. Yamamoto, Phys. Rev. A, {\bf 82}, 013645 (2010).
\bibitem{note1}
The situation that we consider here can be interpreted in two ways. One is that the container is moving with a constant velocity $-\bm{v}\equiv -\hbar\bm{q}/m$. The factor $e^{i\bm{q}\cdot\bm{r}}$ in Eq.~(\ref{eq:Bloch_wave_function_for_condensate_wave_function}) arises from the Galilei transformation. Another is that the condensate is moving with the velocity $\bm{v}$ in the frame of the system at rest. We can not distinguish the above two situations in the thermodynamic limit as long as we focus on excitation branches written by Eq.~(\ref{eq:Bloch_wave_function_for_condensate_wave_function}). See, F. Bloch, Phys. Rev. A, {\bf 7}, 2187 (1973).
\bibitem{supplemental}
See Supplemental Material for details of calculations.
\bibitem{note2}
The authors of Ref.~\onlinecite{Watanabe2012} used the chemical potential as a thermodynamic variable. Therefore, the coexistence phase did not appear in their calculations. 
\bibitem{note3}
In cold atomic gases, LI does not necessarily occur at sufficiently low temperature because emissions of the energy by the condensate are prohibited due to the lack of thermal component that receives the emitted energy. Therefore, the actual instability line is given by DI line in the cold atomic system at low temperature. See, for example, L. De Sarlo, L. Fallani, J. E. Lye, M. Modugno, R. Saers, C. Fort, and M. Inguscio, Phys. Rev. A, {\bf 72}, 013603 (2005); and K. Iigaya, S. Konabe, I. Danshita, and T. Nikuni, {\it ibid}, {\bf 74}, 053611 (2006).
\bibitem{Son2005}
D. T. Son, Phys. Rev. Lett. {\bf 94}, 175301 (2005), Jinwu Ye, Eur. Phys. Lett. {\bf 82}, 16001 (2008). 
\bibitem{Kunimi2011}
M. Kunimi, Y. Nagai, and Y. Kato, Phys. Rev. B, {\bf 84}, 094521 (2011).
\bibitem{Pitaevskii1984}
L. P. Pitaevskii, JETP Lett. {\bf 39}, 511 (1984), Y. Pomeau and S. Rica, Phys. Rev. Lett. {\bf 71}, 247 (1993).
\bibitem{Ancilotto2005}
F. Ancilotto, F. Dalfovo, L. P. Pitaevskii, and F. Toigo, Phys. Rev. B {\bf 71}, 104530 (2005).
\bibitem{Anagama2012}
G. Anagama, M. Kunimi, and Y. Kato, (unpublished).
\end{thebibliography}

\begin{thebibliography}{99}
\bibitem{Arahata2009}
E. Arahata and T. Nikuni, Phys. Rev. A, {\bf 79}, 063606 (2009).
\bibitem{Sepulveda2010}
N. Sep\'{u}lveda, C. Josserand, and S. Rica, Eur. Phys. J. B {\bf 78}, 439 (2010).
\bibitem{note1}
We note that the sign of $\bm{G}$ in eq.~(\ref{eq:expansion_of_v_two_dimension}) will be changed by the definition of $v$ in eq.~(\ref{eq:linear_stability_analysis}).
\end{thebibliography}

%%%%%%%%%%%%%%%%%%%%%%%%%%%%%%%%%%%%%%%%%%%%%%%%%%%%%%%%%%%%%

\newpage
%%%%%%%%%%%%%%%%%%%%%%%%%%%%%%%%%%%%%%%%%%%%%
\begin{widetext}
\section*{Supplemental Material}
\section{Numerical method for the GP equation}

%%%%%%%%%%%%%%%%%%%%%%%%%%%%%%%%%%%%%%%%%%%%%
%%%%%%%%%%%%%%%%%%%%%%%%%%%%%%%%%%%%%%%%%%%%%
In this appendix, we show the numerical method for the Gross-Pitaevskii(GP) equation. A similar method was used in Ref.~\onlinecite{Arahata2009}. In the following, we consider two-dimensional system. 

The time-dependent GP equation is given by
\begin{eqnarray}
i\hbar\frac{\partial}{\partial t}\Psi(\bm{r}, t)&=&-\frac{\hbar^2}{2m}\nabla^2\Psi(\bm{r}, t)+\int d\bm{r}'V(\bm{r}-\bm{r}')|\Psi(\bm{r}', t)|^2\Psi(\bm{r}, t).\label{eq:GP_equation_two_dimenisional_system}
\end{eqnarray}
Substituting $\Psi(\bm{r}, t)=e^{-i\mu t/\hbar}\Psi(\bm{r})$ into Eq.~(\ref{eq:GP_equation_two_dimenisional_system}), we obtain the time-independent GP equation
\begin{eqnarray}
-\frac{\hbar^2}{2m}\nabla^2\Psi(\bm{r})+\int d\bm{r}'V(\bm{r}-\bm{r}')|\Psi(\bm{r}')|^2\Psi(\bm{r})=\mu\Psi(\bm{r}),\label{eq:time-independent_GP_equation_two_dimenisional_system}
\end{eqnarray}
where $\mu$ is a chemical potential determined by the total particle number condition
\begin{eqnarray}
N=\int d\bm{r}|\Psi(\bm{r})|^2,
\end{eqnarray}
where $N$ is the total particle number. We assume that the ground state wave function can be written by the Bloch wave function :
\begin{eqnarray}
\Psi(\bm{r})\equiv e^{i\bm{q}\cdot\bm{r}}\phi(\bm{r}),
\end{eqnarray}
where $\phi(\bm{r})$ is a periodic function that satisfies $\phi(\bm{r}+n_1\bm{a}_1+n_2\bm{a}_2)=\phi(\bm{r})$ for arbitrary integers  $n_1$ and $n_2$ and primitive vectors $\bm{a}_1$ and $\bm{a}_2$. From this assumption, we can expand $\Psi(\bm{r})$:
\begin{eqnarray}
\Psi(\bm{r})=e^{i\bm{q}\cdot\bm{r}}\sum_{\bm{G}}C_{\bm{q}+\bm{G}}e^{i\bm{G}\cdot\bm{r}},\label{eq:Bloch_wave_function_expansion_form}
\end{eqnarray}
where $\bm{G}$ is a reciprocal lattice vector and $C_{\bm{q}+\bm{G}}$ is a  expansion coefficient. In the following, we abbreviate $C_{\bm{q}+\bm{G}}$ to $C_{\bm{G}}$. Substituting Eq.~(\ref{eq:Bloch_wave_function_expansion_form}) into the GP equation and multiplying both sides of the resultant equation by $\displaystyle{\int_{\rm u. c.}}d\bm{r}e^{-i\bm{G}\cdot\bm{r}}$, we obtain
\begin{eqnarray}
&&\left[\frac{\hbar^2}{2m}(\bm{q}+\bm{G})^2+n_0\bar{V}(\bm{0})\right]C_{\bm{G}}+\sum_{\Delta\bm{G}\not={\bm{0}}}S_{\Delta\bm{G}}C_{\bm{G}+\Delta\bm{G}}=\mu C_{\bm{G}},\label{eq:GPeq_in_2D_system_for_expansion_coefficient}\\
&&\hspace{7.5em}S_{\Delta\bm{G}}\equiv \bar{V}(\Delta\bm{G})\sum_{\bm{G}'}C^{\ast}_{\bm{G}'+\Delta\bm{G}}C_{\bm{G}'},
\end{eqnarray}
where we introduce the Fourier transform of the two-body interaction
\begin{eqnarray}
\bar{V}(\bm{k})\equiv \int d\bm{r}e^{-i\bm{k}\cdot\bm{r}}V(\bm{r}),
\end{eqnarray}
and we used the orthonormality and the total particle number condition
\begin{eqnarray}
&&\frac{1}{S}\int_{\rm{u. c.}} d\bm{r}e^{i(\bm{G}-\bm{G}')\cdot\bm{r}}=\delta_{\bm{G}, \bm{G}'},\\
&&n_0\equiv \frac{N}{S}=\sum_{\bm{G}}|C_{\bm{G}}|^2,
\end{eqnarray}
where $S$ is the area of the unit cell and $\displaystyle{\int_{\rm u. c.}d\bm{r}}$ denotes the integral over the unit cell and $n_0$ is the mean-particle density. In the case of the soft-core interaction, $\bar{V}(\bm{k})$ is given by\cite{Sepulveda2010}
\begin{eqnarray}
\bar{V}(\bm{k})=2\pi V_0a^2\frac{J_1(ka)}{ka},
\end{eqnarray}
where $J_1(x)$ is a Bessel function. For example, $S$ is given by
\begin{eqnarray}
S&=&\lambda^2\quad (\text{square lattice}),\\
S&=&\frac{\sqrt{3}}{2}\lambda^2\quad (\text{triangular lattice}),
\end{eqnarray}
where $\lambda$ is a lattice constant that is determined by minimizing the total energy per particle:
\begin{eqnarray}
\frac{E}{N}&=&\frac{\hbar^2}{2mN}\int_{\rm u.c.}d\bm{r}|\nabla\Psi(\bm{r})|^2+\frac{1}{2N}\int_{\rm u.c.}d\bm{r}\int d\bm{r}'V(\bm{r}-\bm{r}')|\Psi(\bm{r}')|^2|\Psi(\bm{r})|^2\nonumber \\
&=&\frac{\hbar^2}{2m n_0}\sum_{\bm{G}}(\bm{q}+\bm{G})^2|C_{\bm{G}}|^2+\frac{1}{2n_0}\sum_{\bm{G}_1, \bm{G}_2, \bm{G}_3}\bar{V}(\bm{G}_1-\bm{G}_3)C_{\bm{G}_1+\bm{G}_2-\bm{G}_3}^{\ast}C_{\bm{G}_3}^{\ast}C_{\bm{G}_2}C_{\bm{G}_1}.
\end{eqnarray}

Here, regarding the GP equation~(\ref{eq:GPeq_in_2D_system_for_expansion_coefficient}) as the eigenvalue equation for $\mu$, we solve the GP equation numerically. Since the GP equation is the non-linear equation for $C_{\bm{G}}$, we need to solve it self-consistently. The advantage of this method is to apply the same diagonalization algorithm to solving the Bogoliubov equation described later. The detail of the procedure is as follows:
\begin{itemize}
\item[(i)] Choose appropriate value of $\lambda$.
\item[(ii)]Choose appropriate $\{C_{\bm{G}}\}$ as an initial condition.
\item[(iii)]Substitute $\{C_{\bm{G}}\}$ into $S_{\Delta\bm{G}}$ in the left-hand side of Eq.~(\ref{eq:GPeq_in_2D_system_for_expansion_coefficient}) and diagonalize Eq.~(\ref{eq:GPeq_in_2D_system_for_expansion_coefficient}) numerically.
\item[(iv)]Calculate the total particle energy per particle for each eigenstate.
\item[(v)]Choose $\{C_{\bm{G}}\}$ for the  lowest energy state.
\item[(vi)]Iterate (iii)-(v) until convergence.
\item[(vii)]Choose different value of $\lambda$ to minimize the total energy per particle.
\item[(viii)]Iterate (ii)-(vii) until convergence.
\end{itemize}

%%%%%%%%%%%%%%%%%%%%%%%%%%%%%%%%%%%%%%%%%%%%%
\section{Numerical method for the Bogoliubov equation}
%%%%%%%%%%%%%%%%%%%%%%%%%%%%%%%%%%%%%%%%%%%%%
In this appendix, we show the numerical method for solving the Bogoliubov equation.  Substituting 
\begin{eqnarray}
\Psi(\bm{r}, t)=e^{-i\mu t/\hbar}\left[\Psi(\bm{r})+u(\bm{r})e^{-i\epsilon t/\hbar}-v^{\ast}(\bm{r})e^{i\epsilon t/\hbar}\right]\label{eq:linear_stability_analysis}
\end{eqnarray}
into Eq. (\ref{eq:GP_equation_two_dimenisional_system}) and retaining $u(\bm{r})$ and $v(\bm{r})$ up to $O(u(\bm{r}))$ and $O(v(\bm{r}))$, we obtain the Bogoliubov equation%\cite{Pitaevskii2003}
\begin{eqnarray}
\epsilon u(\bm{r})&=&K u(\bm{r})+\int d\bm{r}'V(\bm{r}-\bm{r}')\left[\Psi^{\ast}(\bm{r}')\Psi(\bm{r})u(\bm{r}')-\Psi(\bm{r}')\Psi(\bm{r})v(\bm{r}')\right],\label{eq:Bogoliubov_two_dimensional_for_u}\\
\epsilon v(\bm{r})&=&-Kv(\bm{r})-\int d\bm{r}'V(\bm{r}-\bm{r}')\left[\Psi(\bm{r}')\Psi^{\ast}(\bm{r})v(\bm{r}')-\Psi^{\ast}(\bm{r}')\Psi^{\ast}(\bm{r})u(\bm{r}')\right],\label{eq:Bogoliubov_two_dimensional_for_v}\\
K&\equiv &-\frac{\hbar^2}{2m}\nabla^2-\mu+\int d\bm{r}'V(\bm{r}-\bm{r}')|\Psi(\bm{r}')|^2,
\end{eqnarray}
Assuming the crystalline order, we can apply the Bloch theorem to the Bogoliubov equation. $u(\bm{r})$ and $v(\bm{r})$ can be expanded by the reciprocal lattice vector\cite{note1} :
\begin{eqnarray}
u_{\bm{k}, n}(\bm{r})&=&e^{i\bm{q}\cdot\bm{r}}\sum_{\bm{G}}A_{\bm{k}+\bm{G}}e^{i(\bm{k}+\bm{G})\cdot\bm{r}},\label{eq:expansion_of_u_two_dimension}\\
v_{\bm{k}, n}(\bm{r})&=&e^{-i\bm{q}\cdot\bm{r}}\sum_{\bm{G}}B_{\bm{k}+\bm{G}}e^{i(\bm{k}-\bm{G})\cdot\bm{r}},\label{eq:expansion_of_v_two_dimension}
\end{eqnarray}
where $\hbar \bm{k}$ is a quasi-momentum of the excitations and $n$ is the band index. In the following, we abbreviate $A_{\bm{k}+\bm{G}}$ and $B_{\bm{k}+\bm{G}}$ to $A_{\bm{G}}$ and $B_{\bm{G}}$, respectively. Substituting Eqs.~(\ref{eq:expansion_of_u_two_dimension}) and (\ref{eq:expansion_of_v_two_dimension}) into the Bogoliubov equation and multiplying both sides of Eqs.~(\ref{eq:Bogoliubov_two_dimensional_for_u}) and (\ref{eq:Bogoliubov_two_dimensional_for_v}) by $\displaystyle{\int_{\rm u. c.} d\bm{r}}e^{-i\bm{q}\cdot\bm{r}}e^{-i\bm{k}\cdot\bm{r}}e^{-i\bm{G}\cdot\bm{r}}$ and $\displaystyle{\int_{\rm u. c.} d\bm{r}}e^{i\bm{q}\cdot\bm{r}}e^{-i\bm{k}\cdot\bm{r}}e^{i\bm{G}\cdot\bm{r}}$, respectively, we obtain the Bogoliubov equation for the expansion coefficients :
\begin{eqnarray}
&&D^+_{\bm{G}}A_{\bm{G}}+\sum_{\Delta\bm{G}\not=\bm{0}}S^+_{\bm{G},\Delta\bm{G}}A_{\bm{G}+\Delta\bm{G}}-\sum_{\Delta\bm{G}}W^+_{\bm{G}, \Delta\bm{G}}B_{\bm{G}+\Delta\bm{G}}=\epsilon_{\bm{k}, n}A_{\bm{G}},\label{eq:Bogoliubov_coefficient_1}\\
&&-D^-_{\bm{G}}B_{\bm{G}}-\sum_{\Delta\bm{G}\not=\bm{0}}S^{-\ast}_{\bm{G},\Delta\bm{G}}B_{\bm{G}+\Delta\bm{G}}+\sum_{\Delta\bm{G}}W^{-\ast}_{\bm{G}, \Delta\bm{G}}A_{\bm{G}+\Delta\bm{G}}=\epsilon_{\bm{k}, n}B_{\bm{G}},\label{eq:Bogoliubov_coefficient_2}\\
&&D_{\bm{G}}^{\pm}\equiv \frac{\hbar^2}{2m}(\bm{q}\pm\bm{k}+\bm{G})^2-\mu+n_0\bar{V}(\bm{0})+\sum_{\bm{G}'}\bar{V}(\pm\bm{k}+\bm{G}-\bm{G}')|C_{\bm{G}'}|^2,\\
&&S_{\bm{G}, \Delta\bm{G}}^{\pm}\equiv \sum_{\bm{G}'}\left[\bar{V}(\Delta\bm{G})+\bar{V}(\pm\bm{k}+\bm{G}-\bm{G}')\right]C^{\ast}_{\bm{G}'+\Delta\bm{G}}C_{\bm{G}'},\\
&&W_{\bm{G}, \Delta\bm{G}}^{\pm}\equiv \sum_{\bm{G}'}\bar{V}(\pm\bm{k}+\bm{G}-\bm{G}')C_{2\bm{G}+\Delta\bm{G}-\bm{G}'}C_{\bm{G}'}.
\end{eqnarray}
Substituting the solution of the GP equation into Eqs.~(\ref{eq:Bogoliubov_coefficient_1}) and (\ref{eq:Bogoliubov_coefficient_2}) and diagonalizing these equations, we obtain the excitation spectrum $\epsilon_{\bm{k}, n}$.

\end{widetext}

\begin{acknowledgments}
We thank E. Arahata and D. A. Takahashi for useful comments.
\end{acknowledgments}

\end{document}